\documentclass[preprint,12pt, a4paper]{elsarticle}

\usepackage{amssymb}
\usepackage{hyperref}
\usepackage{color}
\usepackage{colortbl}
\usepackage{graphicx}
\usepackage{listings}
\usepackage{xcolor}
\usepackage{makecell}
\usepackage{xurl}

\definecolor{codegreen}{rgb}{0,0.6,0}
\definecolor{codegray}{rgb}{0.5,0.5,0.5}
\definecolor{codepurple}{rgb}{0.58,0,0.82}
\definecolor{backcolour}{rgb}{0.97,0.97,0.97}

\lstdefinestyle{mystyle}{
    backgroundcolor=\color{backcolour},   
    commentstyle=\color{codegreen},
    keywordstyle=\color{magenta},
    numberstyle=\tiny\color{codegray},
    stringstyle=\color{codepurple},
    basicstyle=\ttfamily\footnotesize,
    breakatwhitespace=false,         
    breaklines=true,                 
    captionpos=b,                    
    keepspaces=true,                 
    numbers=left,                    
    numbersep=5pt,                  
    showspaces=false,                
    showstringspaces=false,
    showtabs=false,                  
    tabsize=2
}

\graphicspath{ {./images/} }

\journal{Software Impacts}

\begin{document}

\begin{frontmatter}



\definecolor{LightCyan}{rgb}{0.88,1,1}

\title{Niimpy: a toolbox for behavioral data analysis}


\author{A. Ikäheimonen, A.M. Triana, N. Luong, A. Ziaei, J. Rantaharju, R. Darst, T. Aledavood}

\address{Department of Computer Science, Aalto university, Tietotekniikan laitos, P.O.Box 15400, FI-00076 AALTO}

\begin{abstract}
Behavioral studies using personal digital devices typically produce rich longitudinal datasets of mixed data types. These data provide information about the behavior of users of these devices in real-time and in the users' natural environments. Analyzing the data requires multidisciplinary expertise and dedicated software. Currently, no generalizable, device-agnostic, freely available software exists within Python scientific computing ecosystem to preprocess and analyze such data. 
This paper introduces a Python package, Niimpy, for analyzing digital behavioral data. The Niimpy toolbox is a user-friendly open-source package that can quickly be expanded and adapted to specific research requirements. The toolbox facilitates the analysis phase by offering tools for preprocessing, extracting features, and exploring the data. It also aims to educate the user on behavioral data analysis and promotes open science practices. Over time, Niimpy will expand with extra data analysis features developed by the core group, new users, and developers. Niimpy can help the fast-growing number of researchers with diverse backgrounds who collect data from personal and consumer digital devices to systematically and efficiently analyze the data and extract useful information. This novel information is vital for answering research questions in various fields, from medicine to psychology, sociology, and others.
\end{abstract}


\end{frontmatter}


\noindent

\noindent
\section{Motivation and significance}

Digital behavioral studies aim to quantify human behavior continuously in natural living environments using data from personal digital devices (e.g., smartphones and fitness trackers) and online social media platforms the user interacts with \cite{marsch_digital_2021, onnela_opportunities_2021, insel_digital_2017}. Recently, there has been an increasing scientific interest in digital behavioral studies for unobtrusive human behavior monitoring \cite{barnett_relapse_2018, huckins_mental_2020, info:doi/10.2196/mhealth.9472, aledavood_data_2017}. 

\begin{figure}[ht]
    \centering
    \includegraphics[width=\textwidth]{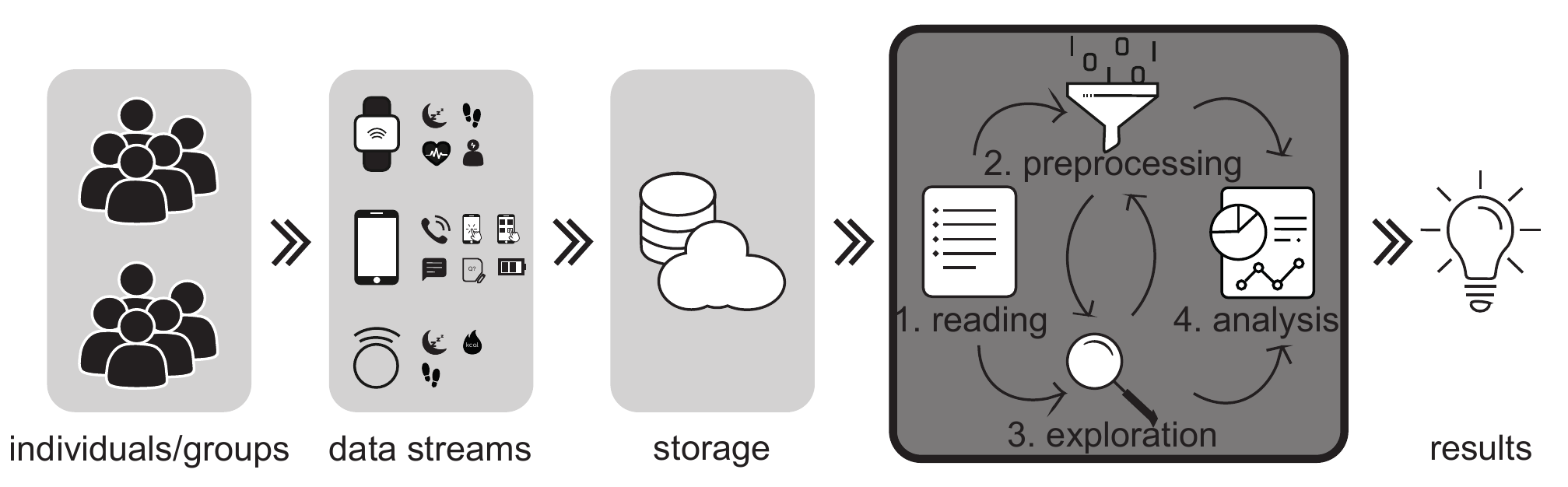}
    \caption{Schematic of a typical digital behavioral study workflow. Generally, individuals or groups volunteer to gather data. The data streams are collected via devices that use different sensors. The data streams are then stored locally or in the cloud. Next, the data is ready to be analyzed. Niimpy provides tools for this analysis phase via four layers (highlighted in dark gray). Users can (i) read the data and then opt to (ii) preprocess or (iii) explore them. While in these layers, users can switch from preprocessing to exploration as needed and once they are ready, start the (iv) analysis via Niimpy to obtain the requested results.}
    \label{fig:Features}
\end{figure}

Typically, digital behavioral studies produce large, heterogeneous, rich data sets of mixed data types. Multiple stages are required to go from the data produced within such studies to meaningful and useful information. A typical digital behavioral study analysis workflow consists of data collection, storage, and analysis phases. Figure \ref{fig:Features} expands on the details. Data preprocessing and feature extraction are crucial tasks for analyzing the data, yet they often have to be re-implemented for each study. Recycling the implemented tools is challenging due to differences in data types and structures across projects. The lack of established data analysis methods and reusable open-source software form significant barriers for new research \cite{vega_reproducible_2021, bent_digital_2021, onnela_beiwe_2021}. Additionally, the lack of reusable methods leads to study results that are not necessarily reproducible and comparable.

The existing software solutions for digital behavioral studies can be categorized by the functionality into three categories: 1) data collection platforms (such as AWARE \cite{ferreira_aware_2015} and BEIWE \cite{torous_new_2016}), 2) data analysis frameworks (e.g., Forest \cite{noauthor_home_nodate} and Rapids \cite{vega_reproducible_2021}), and 3) platforms dedicated for study participants and clinicians (e.g., The Digital Biomarker Discovery Pipeline(DBDP) \cite{bent_digital_2021}, HOPES \cite{wang_hopes_2021}). Some solutions focus on certain functionality, while some encompass all three. 
Among the data analysis frameworks, existing software comes with various limitations; the software may not contain a complete suite for data analysis, may be outdated, not actively developed or maintained, tied to some specific data collection platform or device, or is not openly available for researchers.   

To address these problems, we propose Niimpy: a toolbox for behavioral data analysis. Niimpy is a user-friendly, open-source Python package. It can quickly expand and adapt to specific research questions and workflows and integrates seamlessly into the existing Python scientific computing ecosystem. Niimpy is accompanied by comprehensive documentation and examples using real data, facilitating the implementation of the toolbox. The toolbox is not an out-of-the-box software solution but offers a base framework requiring some programming knowledge to use it.

\section{Software description}

The Niimpy toolbox is a Python package dedicated to rich multi-sensor longitudinal behavioral data analysis. Niimpy can preprocess raw sensor (e.g., GPS coordinates) data or work on predefined data summaries (e.g., daily step count). It is designed for small to moderate-sized (order of thousands of participants) studies. Niimpy is built around Pandas \cite{team_pandas-devpandas_2020} and other Scientific Python stack \cite{2020SciPy-NMeth} libraries. The toolbox requires basic Python programming knowledge. Niimpy comes with comprehensive example notebooks which serve as boilerplate templates for users.

\subsection{Design philosophy}

Niimpy provides basic behavioral data analysis operations and is a starting point for implementing new analyses. As a single software tool cannot provide everything required for every type of analysis, Niimpy makes it easy to customize analyses while building on existing work. 

Niimpy can serve as a framework for other tools, and many new straightforward analysis functions are implemented directly in Niimpy. If these add-on analysis functions are generalizable and reusable, we encourage the user to incorporate those into Niimpy. A standard data schema facilitates generalizability, guides data structures, and promotes overall reusability. As part of Niimpy, these are Pandas dataframes and specific standard column names and data types.

Finally, Niimpy uses existing tools as much as possible. The SciPy ecosystem \cite{2020SciPy-NMeth} provides a wide variety of data processing tools, which are used directly. For example, a significant part of data preprocessing can be done using standard Pandas operations, and shortcut functions are not created for this. Instead, shortcuts are created when they can significantly reduce the user's cognitive load or increase the code's readability. Reusing these standard components allows others to begin using Niimpy more quickly and apply the skills learned in Niimpy to other projects.

\subsection{Software architecture}

The Niimpy software architecture is divided into distinct functional layers; 1) reading, 2) preprocessing, 3) exploration, and 4) analysis. Table \ref{table:3} expands on these details.

\noindent

\subsubsection{Data reading}
The reading layer imports the data from files or other sources, converting the input data to Pandas dataframes with a standard format and doing some minimal data type standardization. Niimpy provides importer functions for CSV and sqlite3 databases; however, in many cases, the user will load and convert data to dataframe format themselves. Niimpy requires data to follow a predefined schema. The schema expects data to be in a tabular (relational) format where a row represents an observation, and columns are properties of observations. This layer is not concerned with the type of sensor or sensor-specific data schemas.

\subsubsection{Data preprocessing}
The preprocessing layer is for data cleaning, filtering, transformation, encoding, and feature extraction. The main focus is on feature extraction functions, while we recommend using existing Python functions for preprocessing when possible. Some preprocessing functions are sensor or device-specific (e.g., Polar tracker feature extraction functions), while some apply generally (e.g., location data feature extraction). Preprocessing functions take in Pandas dataframes and return dataframes and may also require specified column names. Niimpy provides a set of ready-made features for each sensor. Users can extract all the features by default or select the desired ones. Furthermore, users may implement their preprocessing functions.

Currently supported sensors and data streams include the Polar fitness tracker, Android and iOS mobile phone sensors (application data, audio, battery, communication, screen), and device agnostic location and survey streams. For details, refer to appendix table \ref{table:4}.

\subsubsection{Data exploration}

The exploration layer produces visual summaries of data and assesses the data quality (e.g., missing data, outliers). The module includes functions for plotting categorical data counts and distributions, individual and group-wise observation counts, time series line plots for visualizing trends, cyclicity and anomalies, punchcard charts for comprehensive surveys, and visualizing missing data. All the functions are implemented using Plotly Python Open Source Library \cite{inc_collaborative_2015}. Plotly enables interactive inspection of different aspects of the data (e.g., specific time range). For more details about exploration module functions, refer to appendix table \ref{table:5}.

\subsubsection{Data analysis}

The analysis layer has functions for modeling data and statistical inference. The models and inference are applied to the data to identify relationships (e.g., correlation and association) among the features. This layer will be continuously expanded in the future.

\subsection{Unit tests}

A set of unit tests accompanies each Niimpy module to ensure it works as intended. The general guideline is to have unit tests for each function with which the toolbox users interact. The toolbox comes with synthetically created and openly available sample data sets for unit tests. More details about the sample data are found on the GitHub repository \cite{noauthor_niimpy_nodate}. Currently, the code test coverage is above $85\%$.

\subsection{Expanding the toolbox} 

Niimpy can serve as a base to develop new analysis methods by leveraging the lower functional layers. The toolbox is intended to be expanded according to different research needs. We encourage users to contribute reusable and well-generalizable functions, which are accompanied by test functions to the toolbox. The toolbox structure is fully modular; therefore, adding a new feature is straightforward. New features should be added via GitHub pull requests, following the instructions provided in the project code repository. The instructions include required data schema, basic design guidelines, unit test requirements, instructions for documentation and demonstrative notebooks, and instructions for adding example datasets.

\section{Illustrative Examples}

\begin{figure}[ht]
    \centering
    \includegraphics[width=\textwidth]{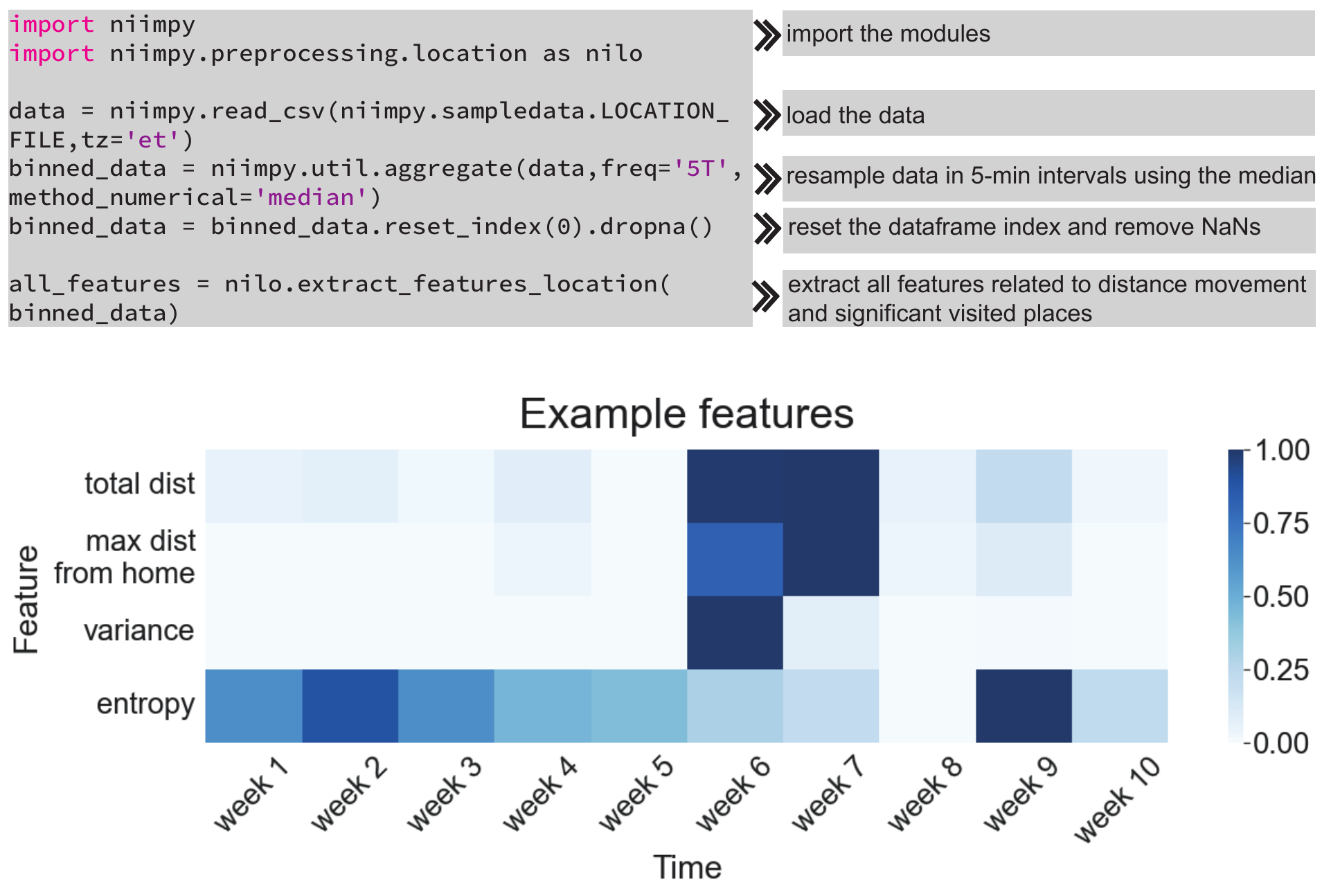}
    \caption{Niimpy usage example: a code example is given to illustrate a Niimpy use case. Here, each line serves to import the modules, prepare the data, and extract the desired features from a GPS data example. Different Niimpy layers are used in this example. First, the data is loaded (reading layer), and then the data is preprocessed (preprocessing layer) and visualized (exploration layer) to understand the movement patterns of a study participant. The figure implies that during weeks 6 and 7, the person has traveled more than during the other weeks.}
    \label{fig:visualization}
\end{figure}

This section provides an illustrative example of extracting location features from GPS data. The features carry useful information for analyzing people's behavioral patterns and changes in their behavior. The example uses a subset of the StudentLife dataset~\cite{noauthor_studentlife_2021}, containing GPS coordinates for one study participant. Figure~\ref{fig:visualization} represents a set of extracted location features from the above example. For more examples, refer to Niimpy's documentation~\cite{noauthor_niimpy_nodate}. 

\section{Impact}

Quantifying digital behavioral data yields information about the study participants' behavioral patterns, changes in patterns, and differences between groups. This information is beneficial for predicting future changes in a person's well-being or clinical conditions. It may also yield new insights into theoretical models of human behavior. However, the key to producing reliable and validated results is to be able to run similar studies in different places and different populations. The studies need to be reproducible, both in terms of study protocols as well as data analysis. Niimpy can facilitate this by making the data analysis workflow consistent from one study to another and making the data analysis more accessible for a wider group of researchers. The resulting information can be used for well-being applications encouraging improved health behavior or may help develop novel, efficient healthcare solutions. 

The Niimpy toolbox helps researchers to analyze digital behavioral data. For this data, preprocessing and feature extraction are critical tasks and have the highest barrier to entry in the analysis. Thus, preprocessing functionalities are the most prominent feature of the toolbox, setting it apart from other data analysis software tools. Niimpy provides tools and comprehensive examples of how to conduct the analysis. 

Niimpy's development was motivated by the needs of the Mobile Monitoring of Mood (MoMo-Mood) pilot \cite{triana_mobile_2020} and the main study, which collected various types of data from different devices and different groups of patients with mental disorders as well as a control group. The toolbox is actively used for these and other similar studies and is consciously extended. In the future, we will continue to develop actively and add new features to the toolbox. While analyzing data sets acquired from existing and new digital behavior studies, we will incorporate a layer for analysis functions into the toolbox. Further, the toolbox is planned to be used for educational purposes in courses and workshops covering the topic of \emph{digital health and human behavior} at Aalto University, Finland, and possibly internationally in the future.

\section{Conclusion}

We have released a Python package, Niimpy toolbox, for digital behavioral data analysis. The toolbox is intended for data scientists and provides data loading, preprocessing, feature extraction, and visualization tools. Niimpy includes comprehensive documentation and examples covering toolbox functionality and educating the user about digital behavioral studies.
The toolbox contributes to the scientific community by facilitating digital behavioral data preprocessing and analysis. Niimpy offers an adaptable data analysis framework enabling replicable and transparent results. 
As the toolbox is still under development, more advanced analysis features will be incorporated into it in the future. Our work aims to promote open science; thus, we encourage researchers to adopt the toolbox and contribute to it with new analysis features.  

\section*{Acknowledgements}
\label{acknowledgements}

We thank professor Jari Saramäki for providing valuable feedback. We also thank Aalto Science-IT for providing computational resources and Aalto Research Software Engineers for their support. We thank Anna Hakala for their help with the project in its early stages. TA acknowledges the support of professor Erkki Isometsä and her other collaborators in the MoMo-Mood project, which has motivated the creation of the Niimpy toolbox. 


\bibliographystyle{elsarticle-num} 
\bibliography{references}

\section*{Required Metadata}
\label{required_metadata}

\section*{Current code version}
\label{current_code_version}

\begin{table}[!ht]
\begin{tabular}{|l|p{6.5cm}|p{6.5cm}|}
\hline
\textbf{Nr.} & \textbf{Code metadata description} & \textbf{Please fill in this column} \\
\hline
C1 & Current code version & 1.0 \\
\hline
C2 & Permanent link to code/repository used for this code version & \url{https://github.com/digitraceslab/niimpy} \\
\hline
C3  & Permanent link to Reproducible Capsule & -- \\
\hline
C4 & Legal Code License   & MIT licence \\
\hline
C5 & Code versioning system used & Git \\
\hline
C6 & Software code languages, tools, and services used & Python \\
\hline
C7 & Compilation requirements, operating environments \& dependencies &  \url{https://github.com/digitraceslab/niimpy/blob/master/requirements-dev.txt} \\
\hline
C8 & If available Link to developer documentation/manual & \url{https://niimpy.readthedocs.io/en/latest/} \\
\hline
C9 & Support email for questions & talayeh.aledavood@aalto.fi \\
\hline
\end{tabular}
\caption{Code metadata(mandatory)}
\label{table:1} 
\end{table}
\newpage

\section*{Current executable software version}
\label{current_executable_software_version}

\begin{table}[!ht]
\begin{tabular}{|l|p{6.5cm}|p{6.5cm}|}
\hline
\textbf{Nr.} & \textbf{(Executable) software metadata description} & \textbf{Please fill in this column} \\
\hline
S1 & Current software version & 1.0 \\
\hline
S2 & Permanent link to executables of this version  &  \url{https://pypi.org/project/niimpy/} \\
\hline
S3  & Permanent link to Reproducible Capsule & -- \\
\hline
S4 & Legal Software License & MIT License \\
\hline
S5 & Computing platforms/Operating Systems & Any capable of running Python \\
\hline
S6 & Installation requirements \& dependencies & Python and SciPy stack packages \\
\hline
S7 & If available, link to user manual - if formally published include a reference to the publication in the reference list & \url{https://niimpy.readthedocs.io/} \\
\hline
S8 & Support email for questions & talayeh.aledavood@aalto.fi \\
\hline
\end{tabular}
\caption{Software metadata (optional)}
\label{table:2} 
\end{table}

\newpage


\section{Appendix\label{Appendix_A}}

\subsection{Toolbox layers by functionality}

\definecolor{Gray}{gray}{0.9}
\begin{table}[!h]
\begin{tabular}{|l|p{13cm}|}
    \hline
    \textbf{Layer Name} & \textbf{Functionality} \\
    \hline
     Reading & Read data from stream or database into a dataframe \\
    \hline
    Preprocessing & Cleaning, encoding, feature extraction, integration, normalization, reduction, transformation \\
    \hline
    Exploration &  Summary statistics, data visualization, data quality assessment \\
    \hline
    Analysis & Data modeling, statistical inference \\
    \hline
\end{tabular}
\caption{Layer functionality summary. For detailed reference, see Niimpy documentation \cite{noauthor_niimpy_nodate}.}
\label{table:3}
\end{table}

\newpage
\subsection{Toolbox preprocessing features}

\begin{table}[!ht]
\begin{tabular}{|l|p{3cm}|p{9.5cm}|}
    \hline
    \textbf{Sub-module} & \textbf{Device} & \textbf{Features} \\
    \hline
    Fitness tracker & Polar device / Polar API v3 & Daily step count: mean, mean standard deviation, min, max, distribution per hour \\
    \hline
    Audio & Aware & \textit{Time window based features}: count silent, 
    count speech, count loud, min freq, max freq, mean freq, 
    median freq, std freq, min db, max db, mean db, 
    median db, std db \\
    \hline
    Application & Aware / Device Agnostics & Count, duration\\
    \hline
    Battery & Aware & shutdown event timestamp, datapoint occurence, datapoint gaps, battery charge difference \\
    \hline
    Communication & Aware / Device Agnostics & total call duration, mean call duration, median call duration, call duration std, call count, outgoing incoming call ratio, sms count \\
    \hline
    Location & Device Agnostic & \textit{Distance based features}: total distance, variance, log variance, average speed, speed variance, max speed, location bin count. \textit{Significant place related features}: static point count, moving point count, static bin count, max distance from home, number of significant places, number of rarely visited places, number of transitios between significant places, bin count in the top1, top2, top3, top4, and top5 cluster, normalized entropy.  \\
    \hline
    Screen & Aware & screen off timestamp, screen event count,
    screen event duration, min screen event duration, max screen event duration, median screen event duration, mean screen event duration, screen event duration std, screen first unlock timestamp \\
    \hline
    Survey & Device Agnostic & \makecell{Survey score summation: min, max, mean*, std* \\ (\textasteriskcentered : for applicable data)} \\
    \hline
\end{tabular}
\caption{Preprocessing sub-module feature summary table. For detailed reference, see Niimpy documentation \cite{noauthor_niimpy_nodate}.}
\label{table:4}
\end{table}

\newpage 

\subsection{Exploration module description}

\begin{table}[!ht]
\begin{tabular}{|p{2.9cm}|p{2.4cm}|p{2.6cm}|l|}
    \hline
    \textbf{Sub-module} &  \textbf{Data Type} & \textbf{Visualization type} & \textbf{Usage} \\
    \hline
    Categorical plot & Categorical & Barplot & Observation counts and distributions \\
    \hline
    Count plot & Categorical* / Numerical & Barplot / Boxplot & Observation counts and distributions\\
    \hline
    Lineplot & Numerical &  Lineplot & Trend, cyclicity, patterns \\
    \hline
    Punchcard & Categorical* / Numerical & Heatmap & Temporal patterns of counts or values \\
    \hline
    Missingness & Categorical* / Numerical & Barplot / Heatmap & Missing data patterns \\
    \hline
\end{tabular}

\caption{Exploration module function summary. For detailed reference, see Niimpy documentation \cite{noauthor_niimpy_nodate}.}
\label{table:5}
\end{table}

\end{document}